\RequirePackage{ifpdf}
\ifpdf % We are running pdfTeX in pdf mode
\documentclass[pdftex]{sigma}
\else
\documentclass{sigma}
\fi

\def\stackreb#1#2{\ \mathrel{\mathop{#1}\limits_{#2}}}

\newcommand{\Z}{\mathbb Z}

\begin{document}

\allowdisplaybreaks

\renewcommand{\PaperNumber}{059}

\FirstPageHeading

\renewcommand{\thefootnote}{$\star$}

\ShortArticleName{A Critical Phenomenon in Solitonic Ising Chains}

\ArticleName{A Critical Phenomenon in Solitonic Ising Chains\footnote{This paper is a
contribution to the Vadim Kuznetsov Memorial Issue `Integrable
Systems and Related Topics'. The full collection is available at
\href{http://www.emis.de/journals/SIGMA/kuznetsov.html}{http://www.emis.de/journals/SIGMA/kuznetsov.html}}}

\Author{Igor M. LOUTSENKO~$^\dag$ and Vyacheslav P. SPIRIDONOV~$^\ddag$}

\AuthorNameForHeading{I.M. Loutsenko and V.P. Spiridonov}

\Address{$^\dag$~Mathematical Institute, University of Oxford,
24-29 St. Gilles', Oxford, OX1 3LB, UK} 
\EmailD{\href{mailto:loutsenk@maths.ox.ac.uk}{loutsenk@maths.ox.ac.uk}} 

\Address{$^\ddag$~Bogoliubov Laboratory of Theoretical Physics, JINR,
Dubna, Moscow Region, 141980 Russia}
\EmailD{\href{mailto:spiridon@theor.jinr.ru}{spiridon@theor.jinr.ru}}

\ArticleDates{Received December 04, 2006, in f\/inal form April
17, 2007; Published online April 24, 2007}

\Abstract{We discuss a phase transition of the second order taking place
in non-local 1D Ising chains generated by specif\/ic
inf\/inite soliton solutions of the KdV and BKP equations.}

\Keywords{Ising chain; solitons; phase transition}

\Classification{70H06; 82B20}

\begin{flushright}
\it To the memory of Vadim B. Kuznetsov
\end{flushright}

\section[The Korteweg-de Vries solitonic spin chain]{The Korteweg--de Vries solitonic spin chain}

In a series of papers \cite{LS1,LS2}, we described a direct relation
between soliton solutions of integrable hierarchies and lattice gas
systems (e.g., Coulomb gases on two dimensional lattices). The latter
models can be reformulated also as some Ising spin systems with a non-local
exchange. In particular, the grand partition functions of
specif\/ic $N$-site Ising chains for some f\/ixed values of
the temperature were shown to coincide with the tau-functions of
$N$-soliton solutions of the Korteweg--de Vries (KdV)
and Kadomtsev--Petviashvili (KP) equations.

We would like to complete here the consideration of \cite{LS1}
and investigate a critical phenomenon appearing in these
models in the zero temperature limit. The
$N$-soliton solution of the KdV equation $u_t+u_{xxx}-6uu_x=0$
has the form \cite{AS,MS}
\begin{gather*}
u(x, t)=- 2\partial_x^2 \log \tau_N(x, t),
%\label{pottau}
\end{gather*}
where $\tau_N$ is the determinant of a $N\times N$ matrix $M$,
\begin{gather*}
\tau_N=\det M, \qquad M_{ij}=\delta_{ij}+ {2\sqrt{k_i k_j}
\over k_i+k_j} e^{(\theta_i+\theta_j)/2},
%\label{phase}
\\
\theta_i=k_ix-k_i^3 t +\theta_i^{(0)}, \qquad i, j=1, 2, \dots, N.
\nonumber
\end{gather*}
The parameters $k_i$ describe amplitudes of solitons,
$\theta_i^{(0)}/k_i$ are the zero time
phases of solitons, and $k_i^2$ are their velocities.
The tau-function $\tau_N$ admits the following Hirota type representation~\cite{Hir}:
\begin{gather}
\tau_N = \sum_{\mu_i=0,1} \exp \left( \sum_{1\leq i<j \le N } A_{ij}
\mu_i \mu_j + \sum_{i=1}^N \theta_i\mu_i \right),
 \label{N_soliton}
 \end{gather}
where the soliton phase shifts $A_{ij}$ are expressed in terms of
the spectral variables $k_i$ as
\begin{gather*}
e^{A_{ij}}={ (k_i-k_j)^2 \over  (k_i+k_j)^2 }.  %\label{KDV_phase}
\end{gather*}
As remarked in \cite{LS1} for $\theta_i=\theta^{(0)}$,
this $\tau_N$ def\/ines the grand partition function of a lattice gas model
with the chemical potential $\theta^{(0)}$ and $\mu_i$ being
the f\/illing factors of the lattice sites by molecules. The constants
$A_{ij}$ describe interaction energy of the molecules.

Substituting in \eqref{N_soliton}
$\mu_i=(\sigma_i+1)/2$, where $\sigma_i=\pm 1$ are other discrete variables,
one can pass from the lattice gases to Ising spin chains \cite{bax}:
\begin{gather*}
\tau_N=e^{\Phi} Z_N, \qquad \Phi=\frac{1}{4}\sum_{i<j} A_{ij}
+ \frac{1}{2}\sum_{j=1}^N \theta_j,
%\label{tauising}
\end{gather*}
where
\begin{gather}
Z_N=\sum_{\sigma_i=\pm1} e^{- \beta E}, \qquad \beta=\frac{1}{kT},
\label{Zising}\\ 
E =\sum_{1\leq i<j \leq N} J_{ij}\sigma_i\sigma_j -\sum_{i=1}^N H_i\sigma_i.
\label{energy}
\end{gather}
Here, $J_{ij}$ are  the
exchange constants, $H_i$ is an external magnetic
f\/ield, $T$ is the temperature, and $k$ is the Boltzmann constant:
\begin{gather*}
\beta J_{ij} =-\frac{1}{4}  A_{ij},
\qquad
\beta H_i=\frac{1}{2}\theta_i +\frac{1}{4}\sum_{j=1, i\neq j}^N A_{ij}.
%\label{identif}
\end{gather*}

$\tau_N$ coincides thus with the partition function of a
one-dimensional Ising chain with the specif\/ic non-local exchange.
A similar situation holds for the KP hierarchy and some other partial
dif\/ferential or dif\/ference nonlinear integrable equations.

From the thermodynamic point of view, it is interesting to understand the $N\to\infty$
behaviour of these ``solitonic" statistical mechanics models. In general
there are inf\/initely many free parameters, and it is
dif\/f\/icult to classify qualitatively dif\/ferent cases. An interesting class
of models is related to the so-called self-similar potentials \cite{spi:universal}.
These potentials are characterized by the $q$-periodicity constraints
$k_{j+M}=qk_j$ and $\theta_{j+M}^{(0)}=\theta_{j}^{(0)}$, where $q$ is
an arbitrary parameter, $0<q<1$, and $M$ is a positive integer.
Exchange constants satisfy in this case the constraints $J_{i+M, j+M}=J_{ij}$.
For $M=1$, this translational invariance takes the simplest form $J_{ij}=J(i-j)$
with $k_i$ forming one geometric progression, $k_i=k_1q^{i-1}$, $q=e^{-2\alpha},$ where $k_1$ and $ \alpha>0$ are free parameters. More precisely,
\begin{gather*}
A_{ij}=2\log |\tanh\alpha (i-j)|.
%\label{interactions} 
\end{gather*}

The KdV coordinate $x$ and time $t$ describe a part of the magnetic f\/ield $H_i$
decaying exponentially fast for $i\to\infty$ because $q<1$.
Only the values of constants $\theta_i^{(0)}$ are therefore relevant for the
leading asymptotics of the partition function in the $N\to\infty$ thermodynamic
limit. Neglecting this $(x,t)$-dependence, we come to the constraint
$H_{i+M}=H_i$, which is just the homogeneity condition for $M=1$.
In principle, it is possible to compute the $N\to\infty$  asymptotics for
the partition function for arbitrary $M$-periodic magnetic f\/ields,
but in \cite{LS1} only the $M=1,2$ cases were considered.

Since $0<|\tanh \alpha(i-j)|<1$, we have $J_{ij} = -A_{ij}/4\beta > 0$,
which corresponds to an antiferromagnetic Ising chain (a similar picture
holds for $M>1$). Although we have a long distance interaction,
its intensity falls of\/f exponentially fast, and the absence of phase
transitions in such systems at non-zero temperatures is well known
\cite{eva}.  It appears, however, that
there exists a special limit leading to a nontrivial critical phenomenon
in this model. Indeed, we consider the limit $\alpha\to 0^+$ or $q\to 1^-$.
The phase shifts $A_{ij}\propto J_{ij}/kT$ then diverge. We can take
nevertheless as the true exchange constants
$J_{ij}^{\rm ren}=J_{ij}(q^{-1}{-}q)$ and as the true temperature
$kT_{\rm ren}=kT(q^{-1}{-}q)$.  For the self-consistency of the temperature def\/inition,
we should renormalize
the magnetic f\/ield as well, $H\propto h/(q^{-1}-q)$, and assume that $h$ is f\/inite.

As a result, the interaction energy of any
spin ``in the bulk" with all others,
\[
E_i=\sum_{j=-\infty,\neq i}^{\infty}
J_{ij}^{\rm ren}\sigma_j,
\]
is f\/inite for $q\to 1^-$ (or $\alpha\to 0^+$). Indeed, the maximal value of this
interaction energy is
\begin{gather*}
E_{\max}=\lim_{\alpha\to0}4\alpha
\sum_{j=-\infty,\neq 0}^\infty J_{0j}=-\frac{2}{\beta} \lim_{\alpha\to0}\alpha
\sum_{j=-\infty,\neq 0}^\infty\log\left|\frac{1-q^j}{1+q^j}\right|
=-\frac{4}{\beta} \lim_{\alpha\to0}\alpha\log\frac{(q;q)_\infty}{(-q;q)_\infty}.
\end{gather*}
We use the notation $(a;q)_\infty=\prod\limits_{k=0}^\infty(1-aq^k)$ and
$(a_1,\ldots, a_m;q)_\infty=\prod\limits_{j=1}^m(a_j;q)_\infty$.

For the following considerations, we need theta functions \cite{aar:special}
\begin{gather*}
 \theta_1(\nu,q)=-i\sum_{n\in\Z}(-1)^nq^{(n+1/2)^2}e^{(2n+1)i\nu}
=iq^{1/4}e^{-i\nu}\big(q^2,e^{2i\nu},q^2e^{-2i\nu};q^2\big)_\infty,
\\ 
\theta_2(\nu,q)=\sum_{n\in\Z}q^{(n+1/2)^2}e^{(2n+1)i\nu}
=q^{1/4}e^{-i\nu}\big(q^2,-e^{2i\nu},-q^2e^{-2i\nu};q^2\big)_\infty,
\\ 
\theta_3(\nu,q)=\sum_{n\in\Z}q^{n^2}e^{2ni\nu}
=\big(q^2,-qe^{2i\nu},-qe^{-2i\nu};q^2\big)_\infty,
\\ 
\theta_4(\nu,q)=\sum_{n\in\Z}(-1)^nq^{n^2}e^{2ni\nu}
=\big(q^2,qe^{2i\nu},qe^{-2i\nu};q^2\big)_\infty,
\end{gather*}
where $q=e^{\pi i\tau}$, $\text{Im}(\tau)>0$,
and their modular transformations
\begin{alignat*}{3}
& \theta_1(\nu/\tau,\tilde q)=-i\sqrt{-i\tau}e^{i\nu^2/\pi\tau}
\theta_1(\nu,q),
\qquad &&
\theta_2(\nu/\tau,\tilde q)= \sqrt{-i\tau}e^{i\nu^2/\pi\tau}\theta_4(\nu,q),&
\\ 
&\theta_3(\nu/\tau,\tilde q)=\sqrt{-i\tau}e^{i\nu^2/\pi\tau}\theta_3(\nu,q),
\qquad&&
\theta_4(\nu/\tau,\tilde q)=\sqrt{-i\tau}e^{i\nu^2/\pi\tau}\theta_2(\nu,q),&
\end{alignat*}
where $\tilde q=e^{-\pi i/\tau}$ and $\sqrt{-i\tau}$ is positive for purely
imaginary $\tau$.
Using these formulas, we obtain
\[
\frac{(q;q)^2_\infty}{(-q;q)^2_\infty}=\frac{\theta_1'(0,q^{1/2})}
{\theta_2(0,q^{1/2})}=\frac{\theta_1'(0,{\tilde q}^{1/2})}
{(-i\tau)\theta_4(0,{\tilde q}^{1/2})}={\tilde q}^{1/8}\frac{2}{(-i\tau)}
\frac{(\tilde q;\tilde q)_\infty^2}{({\tilde q}^{1/2};\tilde q)_\infty^2},
\]
where $q=e^{2\pi i\tau}=e^{-2\alpha}$ and
$\tilde q=e^{-2\pi i/\tau}=e^{-2\pi^2/\alpha}$.
As a result,
\[
E_{\max}=\frac{\pi^2}{2\beta}<\infty,
\]
and $E_i\leq E_{\max}$.
The limit $\alpha\to 0$ corresponds thus to an inf\/initely small and inf\/initely
long-range nonlocal interaction model at a low value (zero) ef\/fective temperature.

There are some other interesting limits. For instance,
for $q \to 0$ and f\/inite $h$, we obtain the high temperature nearest
neighbor interaction spin chain, $J_{ij}^{\rm ren}\propto \delta_{i+1,j}$,
$T_{\rm ren}\to\infty$. For f\/inite~$H$, this limit
corresponds to the non-interacting spins. The solitonic interpretation
describes thus only a two-dimensional subspace of parameters $(T, H, q)$.
For f\/ixed $q$, the temperature $T$ is also f\/ixed, and we can set the
 ``KdV temperature'' equal to $\beta=1$.

Using the Wronskian representation for $\tau_N$,
the leading asymptotics of $Z_N$ for $N\to\infty$ was determined in \cite{LS1}
for the $M=1$ translationally invariant model and a
homogeneous magnetic f\/ield. Namely, $Z_{N}\to \exp(-N\beta f_I)$,
where the free energy per site $f_I$ has the form
\begin{gather*}
-\beta f_I(q, H)=\log \frac{2(q^4;q^4)_\infty \cosh \beta H}{(q^2;q^2)_\infty^{1/2}}
+ \frac{1}{4\pi}\int_0^{2\pi}d\nu \log \big(|\rho(\nu)|^2 - q\tanh^2 \beta H\big),
%\label{free}
\\
|\rho(\nu)|^2=
\frac{(q^2e^{i\nu},q^2e^{-i\nu};q^4)_\infty^2}
{(q^4e^{i\nu},q^4e^{-i\nu};q^4)_\infty^2}\;
\frac{1}{4\sin^2(\nu/2)}=
q\frac{\theta_4^2(\nu/2, q^2)}{\theta_1^2(\nu/2, q^2)}.
\end{gather*}
The total magnetization of the lattice takes the form:
\begin{gather}
m(H)=-\frac{\partial f_I}{ \partial H}
= \stackreb{\lim}{N\to\infty} \frac{1}{N} \sum_{i=1}^{N}\langle \sigma_i \rangle
=\left(1-\frac{1}{\pi}
\int_0^{\pi}\frac{d\nu}{1+d(\nu)\cosh^2\beta H }\right)\tanh \beta H,
\label{magkdv}
\end{gather}
where
\[
d(\nu)=\frac{\theta_4^2(\nu, q^2)}{\theta_1^2(\nu, q^2)}-1.
\]
The function $m(H)$ grows monotonically with $H$ and ref\/lects
qualitative predictions of the general theory of 1D systems with the fast
decaying interactions \cite{eva}.
However, the limit $\alpha\to 0$ with the renormalized exchange and
magnetic f\/ield breaks down the corresponding necessary conditions,
and we obtain a non-trivial critical phenomenon.

We substitute in \eqref{magkdv} $ \beta H=h/(q^{-1}-q)$, $h>0,$
and take the limit $\alpha\to 0$. Since
\[
\frac{\theta_4^2(\nu, q^2)}{\theta_1^2(\nu, q^2)}=
-\frac{\theta_2^2(\nu/\tau, {\tilde q}^2)}{\theta_1^2(\nu/\tau,{\tilde q}^2)}=
\frac{(-e^{2i\nu/\tau},-{\tilde q}^4e^{-2i\nu/\tau};{\tilde q}^4)^2_\infty}
{(e^{2i\nu/\tau},{\tilde q}^4e^{-2i\nu/\tau};{\tilde q}^4)^2_\infty},
\]
where $q=e^{\pi i \tau/2}=e^{-2\alpha}$,
$\tilde q=e^{-\pi i/2 \tau}=q^{-\pi^2/8\alpha}$, we have
\begin{gather*}
 \chi(\nu)\equiv \lim_{\alpha\to 0}d(\nu)\cosh^2\frac{  h}{4\alpha}
=\lim_{\alpha\to 0}\left(\frac{(1+e^{2i\nu/\tau})^2(1+e^{2i(\nu-\pi)/\tau})^2}
{(1-e^{2i\nu/\tau})^2(1-e^{2i(\nu-\pi)/\tau})^2}-1\right)\frac{e^{  h/2\alpha}}{4}
\\  \phantom{\chi(\nu)}{}
=\lim_{\alpha\to 0}\left(e^{-2i\nu/\tau}+e^{2i(\nu-\pi)/\tau}\right)
e^{  h/2\alpha}, \qquad 0< \nu <\pi.
\end{gather*}
Substituting this result in \eqref{magkdv} and using
relation $\lim\limits_{\alpha\to0}\tanh (  h/4\alpha)=1$,
we obtain
\[
m(h)=1-\frac{2}{\pi}
\int_0^{\pi/2}\frac{d\nu}{1+\chi(\nu)}.
\]
For $0< \nu<\pi/2$, we have
\[
\chi(\nu)=\begin{cases}
0, & \text{if}\quad   h/\pi<\nu<\pi/2,  \\
\infty, &\text{if}\quad 0<\nu<  h/\pi,
\end{cases}
\]
for $  h< \pi^2/2$, and $\chi(\nu)=\infty$, for $  h>\pi^2/2$.
The f\/inal result therefore can be represented in the form
\[
m(h)=\begin{cases} \displaystyle 1-\frac{2}{\pi}\int_{  h/\pi}^{\pi/2}d\nu
=\frac{2 }{\pi^2} h, & \text{if}\quad   |h|<\pi^2/2,  \\
1, &\text{if}\quad   |h| \geq \pi^2/2.
\end{cases}
\]
We have an obvious point of non-analyticity of $m(h)$ or of the free
energy $f_I(h)$ at the cri\-tical value of the magnetic f\/ield
$h_{\rm crit}=\pi^2/2 $, such that the magnetic susceptibility
$\chi(h)=\beta^{-1}dm(h)/dh$
has a jump at it (i.e., we have the phase transition of the
second order). This is a~typical phenomenon in the systems with
long-range interaction, where the mean f\/ield approxi\-mation gives exact
values for the one-point correlation functions (see, e.g., \cite{bax}).

\section[The mean field approximation]{The mean f\/ield approximation}

In the mean f\/ield theory, one considers a few degrees of freedom
(usually, just one) of a taken system in an ef\/fective mean
f\/ield of the remaining part of the system.
This ef\/fective or mean f\/ield depends itself on the
analysis of the one-body dynamics. As an example, we consider the
general spin chain with energy (\ref{energy}) and
the mean magnetization at the $i$-th site of the lattice
\[
\langle \sigma_i\rangle =\frac{\sum\limits_{\sigma_1,\sigma_2,\ldots}\sigma_ie^{-\beta E}}
{\sum\limits_{\sigma_1,\sigma_2,\ldots}e^{-\beta E}}.
\]
Instead of calculating the above sums, we stick to the $i$-th spin
and evaluate its contribution to the energy as
\begin{gather}
E_i(\sigma_i)=-\sigma_i\tilde H_i,
\label{energy_i}
\end{gather}
where
\begin{gather}
\tilde H_i=-\sum_{j\not=i}J_{ij}\langle \sigma_j\rangle+H_i
\label{H_i}
\end{gather}
is an ef\/fective mean magnetic f\/ield at the $i$-th site created by the external f\/ield
$H_i$ and the rest of the system $\propto\sum\limits_{j\not=i}J_{ij}\langle \sigma_j\rangle $.

In the one body problem (\ref{energy_i}), the conf\/iguration space
consists of two states $\sigma_i=\pm 1$, and therefore
\[
\langle \sigma_i\rangle =\frac{e^{\beta\tilde H_i}
-e^{-\beta\tilde H_i}}{e^{\beta\tilde H_i}+e^{-\beta\tilde H_i}}
=\tanh(\beta \tilde H_i).
\]
Substituting the values of ef\/fective f\/ields (\ref{H_i}) in the
last equation, we obtain a system of transcendental equations for
mean values of all spins
\begin{gather}
\langle \sigma_i\rangle =\tanh\beta\left(-\sum_{j\not=i}J_{ij}\langle \sigma_j\rangle +H_i\right).
\label{mag_i}
\end{gather}
We consider now the translationally invariant system $J_{ij}=J(i-j)$, $H_i=H$.
In such a system, all mean values of the spins are the same in the
thermodynamic limit, $ \langle \sigma_i\rangle =\langle \sigma\rangle $, and from
(\ref{mag_i}), we obtain
\[
\langle \sigma\rangle =\tanh\beta\left(-\sum_{j\not=i}J(i-j)\langle \sigma\rangle +H\right)
=\tanh\beta(-J\langle \sigma\rangle +H), \qquad J=\sum_{j\not=0}J(j).
\]
The solution of this equation is an intersection of graphs of two functions:
$y=x$ and $y=\tanh\beta(-Jx+H)$. We consider two dif\/ferent cases.

1) {\it A ferromagnet in the zero magnetic field}: $J<0$, $H=0$.
Our system of equations has only the trivial solution
$x=y=0$ for $\beta |J|\le 1$ and three solutions $x=y=0$ and $ x=y=\pm m$
for $\beta |J|> 1$ and some $0<m<1$. The latter nontrivial solutions
describe the spontaneous magnetization $m$ at the temperatures smaller than
the critical value $1/\beta_{\rm crit}=|J|$ (the $\tanh(-\beta Jx)$-function
becomes steeper at the origin as $\beta$ increases and starts to intersect the
line $x=y$ in two additional points as its slope exceeds the critical value).

2) {\it An antiferromagnet}, $J>0$. The only solution at $H=0$ is the
trivial solution $x=y=0$. We consider now the zero-temperature limit
$\beta\to\infty$. In this case, the $\tanh$-function transforms to the
sign-function, and we obtain
\begin{gather*}
\left\{\begin{array}{l}
y=x,\\
y=\text{sgn}(-Jx+H),
\end{array}\right.
\quad
\text{sgn}(x)=\left\{\begin{array}{rl}
1, \  & x>0,\\
-1, \  & x<0.
\end{array}\right.
%\label{step}
\end{gather*}
Solving these equations is rather easy. The function $\text{sgn}(x)$ is shifted
by $H/J$ from the origin along the $x$-axis.
When $|H/J|<1$, it intersects with the
line $y=x$ by its vertical part, and the
magnetization equals to $H/J$. When $|H|$ exceeds $J$,
the line $y=x$ intersects with one of the horizontal
branches of $\text{sgn}(x)$, and the magnetization becomes equal to $\pm 1$.
The magnetization is thus a continuous piecewise linear
function of $H$ consisting of three parts: two constant
$\langle \sigma\rangle =\pm 1$ for $|H|>J$ and the linear piece
$\langle \sigma\rangle =H/J$ connecting them through the origin.
In our KdV-solitonic model, we denoted $\beta H=h/(q^{-1}-q)$
and $E_{\max}=(q^{-1}-q)J$. In the limit $q\to 1$, we have
$E_{\max}=\pi^2/2\beta$ and $\langle \sigma\rangle =H/J=2h/\pi^2$,
which leads to the exact value of the critical magnetic f\/ield $h_{\rm crit}=\pi^2/2$.
Such a qualitative behaviour of the system is obvious: when the external
f\/ield exceeds the interaction energy between spins, they all f\/lip
in the f\/ield direction.

The mean f\/ield approximation is known to give exact one-point correlation
functions (e.g., the magnetization) for systems with the long-range
interaction (as our $q\to 1$ limit). It might be non-suitable, however,
for the two point correlators (e.g., $\langle \sigma_i\sigma_j\rangle $).

We consider now the $M$-periodic chain. For general (not
necessarily solitonic) $M$-periodic chain, it is reasonable
to introduce multi-index exchange
$J^{nm}(i-j)$, where $i-j$ is the distance between the cells
and $1\le n,m\le M$ are the respective internal cell indices for the $n$-th and
$m$-th sublattices. In our particular solitonic KdV case, we have
\begin{gather}
\beta J^{nm}(i)=-\frac{1}{2}\log\left|\frac{k_nq^i-k_m}{k_nq^i+k_m}\right|,
\qquad J^{nm}(i)=J^{mn}(-i),
\label{exchange}
\end{gather}
and the energy
\[
E=\sum_{i,j\in Z,\, n\neq m} J^{nm}(i-j)\sigma^{(n)}_i\sigma^{(m)}_j
+\sum_{i,j\in Z,\, i\neq j,\, n}J^{nn}(i-j)\sigma^{(n)}_i\sigma^{(n)}_j
+\sum_{i\in Z,\, n} H^{(n)} \sigma^{(n)}_i.
\]
In the latter sum the magnetic f\/ield is also $M$-periodic,
$H^{(n)}_i=H^{(n)}$, inhomogeneous only inside the cells.
The analysis similar to the $M=1$ case yields from
(\ref{mag_i}) the following system of $M$ equations for
$M$ unknowns $\langle \sigma^{(n)}\rangle$:
\[
\langle \sigma^{(n)}\rangle =\tanh\beta\left(-\sum_{m} J^{nm}\langle \sigma^{(m)}
\rangle +H^{(n)}\right), \qquad n,m=1,\ldots,M.
\]
In these equations,
\[
J^{nn}=\sum_{i\in Z, \neq 0}J^{nn}(i), \qquad J^{nm}=\sum_{i\in Z}J^{nm}(i), \qquad n\neq m.
\]
In the zero-temperature limit, we have
\begin{equation}
\langle \sigma^{(n)}\rangle ={\rm sgn}\left(-\sum_{m} J^{nm}\langle \sigma^{(m)}
\rangle +H^{(n)}\right), \qquad n,m=1,\ldots,M,
\label{magnets}
\end{equation}
where ${\rm sgn}(x)$ is the sign-function.

In the solitonic case (\ref{exchange}), the $M\times M$ matrix
$J=J^{nm}$ is symmetric with the constant diagonal:
\begin{gather}
J^{nn}=-\frac{1}{2\beta}\sum_{i\in Z,\,i\neq0}\log\left|\frac{1-q^i}{1+q^i}\right|\equiv -A,
 \qquad J^{nm}=J^{mn}=-\frac{1}{2\beta}\sum_{i\in Z}\log
\left|\frac{k_nq^i-k_m}{k_nq^i+k_m}\right|.
\label{exchanges}
\end{gather}
For $M=2$, we have $J^{12}=J^{21}\equiv -B$ and
\begin{gather*}
\langle \sigma^{(1)}\rangle =\text{sgn}(A\langle \sigma^{(1)}\rangle
+B\langle \sigma^{(2)}\rangle +H^{(1)}),\\
\langle \sigma^{(2)}\rangle =\text{sgn}(B\langle \sigma^{(1)}\rangle
+A\langle \sigma^{(2)}\rangle +H^{(2)}).
\end{gather*}
If we take the uniform magnetic f\/ield $H^{(1)}=H^{(2)}=H$, these equations become
symmetric in~$\sigma^{(1)}$ and~$\sigma^{(2)}$ and have the solution
$\langle \sigma\rangle =\langle \sigma^{(1)}\rangle =\langle \sigma^{(2)}\rangle $
stemming from one equation
\[
\langle \sigma\rangle =\text{sgn}((A+B)\langle \sigma\rangle +H).
\]
For the completely uniform magnetic f\/ield, there exists thus a solution when the
spins f\/lip simultaneously for both sublattices for suf\/f\/iciently large magnetic f\/ields.

Simultaneous phase transition exists for all sublattices in the uniform f\/ield
$H^{(1)}=\cdots=H^{(M)}=H$, when $\langle \sigma^{(1)}\rangle
=\cdots=\langle \sigma^{(M)}\rangle$ and all spins in all sublattices are aligned
simultaneously for a suf\/f\/iciently large $H$.
As seen from (\ref{magnets}), such a solution exists,  if  $\sum_m J^{nm}$
are equal, which is certainly true for $M=2$ because of the permutational
symmetry. But it may be not so for $M>2$. For instance,
in the solitonic case (\ref{exchanges}) for $M=3$, we have
\begin{gather*}
\sum_m J^{1m}\propto \sum_{i\in Z, \neq 0}\log\left|\frac{1-q^i}{1+q^i}\right|
+\sum_{i\in Z}\log\left|\frac{k_1q^i-k_2}{k_1q^i+k_2}\right|
+\sum_{i\in Z}\log\left|\frac{k_1q^i-k_3}{k_1q^i+k_3}\right|, \\
\sum_m J^{2m}\propto \sum_{i\in Z, \neq 0}\log\left|\frac{1-q^i}{1+q^i}\right|
+\sum_{i\in Z}\log\left|\frac{k_2q^i-k_1}{k_2q^i+k_1}\right|
+\sum_{i\in Z}\log\left|\frac{k_2q^i-k_3}{k_2q^i+k_3}\right|, \\
\sum_m J^{3m}\propto \sum_{i\in Z, \neq 0}\log\left|\frac{1-q^i}{1+q^i}\right|
+\sum_{i\in Z}\log\left|\frac{k_3q^i-k_1}{k_3q^i+k_1}\right|
+\sum_{i\in Z}\log\left|\frac{k_3q^i-k_2}{k_3q^i+k_2}\right|.
\end{gather*}
These three sums are certainly dif\/ferent for $0<q<1$, and in the limit
$\beta\to \infty$ (which we cannot reach within the solitonic interpretation
for $q<1$), the magnetization would become a~piecewise linear function of $H$
of a more complicated form than in the $M=1,2$ cases.
In our model, however, the zero temperature is reached by multiplication of the
above sums by $q^{-1}-q$ and taking the limit $q\to 1^-$ (or $\alpha\to 0^+$).
All three sums become then equal yielding the same magnetization as in
the $M=1$ and $M=2$ cases.

\section{The BKP solitonic spin chain}

Another Ising chain model solved in \cite{LS1} appears from the
multisoliton solution of the KP equation of $B$ type, i.e.\ the
BKP equation~\cite{DJKM}.
The corresponding partition function has the same form~\eqref{Zising},
where the exchange constants are
\begin{gather*}
\beta J_{ij}=-\frac{1}{4}A_{ij}, \qquad e^{A_{ij}}=\frac{(a_i-a_j)(b_i-b_j)(a_i-b_j)(b_i-a_j)}
{(a_i+a_j)(b_i+b_j)(a_i+b_j)(b_i+a_j)}.
%\label{BKP}
\end{gather*}
For $a_i=b_i=k_i/2$, this model coincides with the KdV-inspired model at
the twice lower value of the temperature obtained after the change
$\beta\to 2\beta$.

The translational invariance of this spin chain, $J_{ij}=J(i-j)$, yields
\begin{gather*}
a_i=q^{i-1}, \qquad b_i=bq^{i-1}, \qquad q=e^{-2\alpha},
%\label{BKP_momentums}
\end{gather*}
where we normalize $a_1=1$ and assume that $0<q<1$ as before.
This gives the exchange
\[
\beta J_{ij}=-\frac{1}{4}\log \frac{\tanh^2\alpha(i-j)- (b-1)^2/(b+1)^2}
{\coth^2\alpha(i-j) - (b-1)^2/(b+1)^2},
\]
where the parameter $b$ is restricted to three regions
(because of the $b\to 1/b$ invariance):
either $-1 < b < -q$ (the ferromagnetic chain, $J_{ij}<0$),
or $q< b \leq 1$ or $|b|=1$, $b\neq -1$ (the antiferromagnetic chain,
$J_{ij}>0$).

In the thermodynamic limit $N \to \infty$,
the free energy per site for the homogeneous magnetic f\/ield $H_i=H$
takes the form \cite{LS1}:
\[
-\beta f_I(H)=\frac{1}{4}\log\frac{(q, q, bq, q/b; q)_\infty}
{(-q, -q, -bq, -q/b; q)_\infty} +\frac{1}{4\pi}\int_0^{2\pi}d\nu \log |2\rho(\nu)|,
\]
where
\begin{gather*}
\rho(\nu)=\cosh 2\beta H
+\frac{(-q;q)_\infty^2}{(-e^{i\nu}, -qe^{-i\nu}; q)_\infty}\!
\left(
\frac{(b^{-1}e^{i\nu}, qbe^{-i\nu}; q)_\infty}{(b^{-1}, qb; q)_\infty}
+\frac{(be^{i\nu}, qb^{-1}e^{-i\nu}; q)_\infty}{(b, qb^{-1}; q)_\infty}
\right).\!\!
%\label{free_bkp}
\end{gather*}
Taking the derivative with respect to $H$, we f\/ind the magnetization
\begin{gather}
m(H) =
\left(1-\frac{1}{\pi}\int_0^\pi \frac{d\nu}
{1+d(\nu)\cosh 2\beta H}\right) \tanh 2\beta H,
\label{magbkpgen}
\end{gather}
where
\begin{gather} d(\nu)=
2\frac{\theta_1(\phi/2,q^{1/2})}{\theta_2(0,q^{1/2})}
\frac{\theta_2(\nu, q^{1/2})}
{\theta_1(\nu+\phi/2, q^{1/2})-\theta_1(\nu-\phi/2, q^{1/2})}
\label{d-func}
\end{gather}
with $b=e^{i\phi}$. For real $\phi$ we have $|b|=1$, the choice
$\phi=i\gamma$, $0<\gamma<2\alpha,$ yields $q<b<1$, and for $\phi=\pi+i\gamma$, we have
$-1<b<-q$. The limit $b\to 1$ describes the magnetization
for the ``KdV-spin chain" at the twice lower value of the temperature.
A simple test of this expression consists in the choice $b=-1$ corresponding to
the non-interacting spins, $J_{ij}=0$. In this case $d(\nu)=1$, and we obtain
$m(H)=\tanh\beta H$ as it should be for the free system.

We substitute now $q=e^{-2\alpha}$ and $H=2h/(q^{-1}-q)$ in \eqref{magbkpgen}
and consider the limit $\alpha\to 0^+$. The factor $2$ in front of $h$ was
chosen for coincidence of this model with the the KdV spin chain with
the ef\/fective replacement $\beta\to 2\beta$ (i.e., the twice lower
value of the temperature). We apply the modular transformation
to theta functions in \eqref{d-func} and obtain
\begin{gather*}
 d(\nu)=
2\frac{\theta_1(\phi/2\tau,{\tilde q}^{1/2})}{\theta_4(0,{\tilde q}^{1/2})}
\frac{\theta_2(\nu/\tau,{\tilde q}^{1/2})}
{e^{-i\frac{\nu\phi}{\pi\tau} }\theta_1\left(\frac{\nu+\phi/2}{\tau},{\tilde q}^{1/2}\right)
-e^{i\frac{\nu\phi}{\pi\tau} }
\theta_1\left(\frac{\nu-\phi/2}{\tau},{\tilde q}^{1/2}\right) },
%\label{d-func-m}
\end{gather*}
where $\tau=i\alpha/\pi$ and $\tilde q=e^{-2\pi i/\tau}=e^{-2\pi^2/\alpha}$.
Denoting
\[
\chi(\nu)=\lim_{\alpha\to0}d(\nu)\cosh \frac{4q  h}{1-q^2},
\]
we therefore obtain
\[
\chi(\nu)=\lim_{\alpha\to0}
\frac{\frac{1}{2}e^{-\frac{\pi \phi}{2\alpha}} \big(1-e^{\frac{\pi\phi}{\alpha}}\big)
\big(1-e^{\frac{\pi(2\nu-\pi)}{\alpha}}\big) e^{\frac{  h}{\alpha}} }
{e^{-\frac{\pi\nu}{\alpha}(1+\frac{\phi}{\pi})-\frac{\pi\phi}{2\alpha}}
\big(1-e^{2\pi\frac{\nu+\phi/2}{\alpha}}\big)
-e^{-\frac{\pi\nu}{\alpha}(1-\frac{\phi}{\pi})+\frac{\pi\phi}{2\alpha}}
\big(1-e^{2\pi\frac{\nu-\phi/2}{\alpha}}\big) }.
\]

In the region $\phi=i\gamma,\,0<\gamma<2\alpha,$ we have
\[
\chi(\nu)
=\lim_{\alpha\to0}\frac{\sin(\pi\gamma/2\alpha)
\big(1-e^{\frac{\pi}{\alpha}(2\nu-\pi)}\big)e^{\frac{  h}{\alpha}-\frac{\pi\nu}{\alpha}} }
{2\sin((\pi/2-\nu)\gamma/\alpha)}.
\]
Since $0< \gamma/\alpha<2$, the $\sin$-factors do not inf\/luence
the asymptotic behaviour, and for $0<\nu<\pi/2$, we f\/ind
$\chi(\nu)=0$ for $  h/\pi<\nu<\pi/2$
and $\chi(\nu)=\infty$ for $\nu<  h/\pi$. As a result, we obtain
$m(h)=2  h/\pi^2$ for $|h|<\pi^2/2 $ and
$m(h)=1$ for $|h|\geq\pi^2/2 $. This is the same picture as for the ``KdV-chain",
as it should be because the limit $\alpha\to 0$ assumes
the limit $\gamma\to0$ or $b\to 1$.

In a similar way, for $\phi=\pi+i\gamma$, $0\leq \gamma<2\alpha,$ and $\alpha\to0$,
we f\/ind $b\to -1$, i.e.\ the trivial situation of free spins.
The most interesting behaviour appears in the region $0<\phi<\pi$,
for which we f\/ind
\[
m(h)=1-\frac{2}{\pi}\int_0^{\pi/2}\frac{d\nu}{1+\chi(\nu)},
\]
where
\[
\chi(\nu)=\begin{cases}
0, & \text{if}\quad   h/(\pi-\phi)<\nu,  \\
\infty, &\text{if}\quad \nu<  h/(\pi-\phi).
\end{cases}
\]
As a result,
\[
m(h)=\frac{2  h}{\pi(\pi-\phi)},\qquad \text{if}\qquad |h|<\frac{\pi(\pi-\phi)}{2},
\]
and $m(h)=1$, if $|h|\geq \pi(\pi-\phi)/2 $.  The critical value of
the magnetic f\/ield, for which we have the phase transition,
\[
h_{\rm crit} =\pm \frac{\pi(\pi-\phi)}{2},
\]
depends explicitly on the parameter of
the model $\phi$, and for $\phi=0$, we obtain the previous result.
The Ising spin systems associated with the multisoliton solutions of
integrable nonlinear  equations provide thus the models with phase transitions
already in their simplest one-dimensional spin chain realizations.

It is interesting to analyze consequences of the antiferromagnetic
nature of the exchange. We take for this the KdV-inspired Ising chain
and apply dif\/ferent magnetic f\/ields to the odd, $H_1$,  and even,
$H_2$, sites. The corresponding magnetization for the odd sites sublattice
was derived in \cite{LS1}:
\begin{gather*}
 m_{\rm odd}(H_1,H_2)=-2\frac{df_I}{dH_1}=\lim_{p\to\infty}
\frac{1}{p} \sum_{j=1}^p  \langle\sigma_{2j-1}\rangle
\\ \phantom{m_{\rm odd}(H_1,H_2)}{}
=\tanh\beta H_1-\frac{\tanh\beta H_2}{\cosh^2\beta H_1}
\frac{1}{\pi}\int_0^\pi\frac{d\nu}
{\frac{\theta_4^2(\nu,q^2)}{\theta_1^2(\nu,q^2)}-\tanh\beta H_1\tanh\beta H_2}.
\end{gather*}
The magnetization for the even sites sublattice is obtained after permuting
$H_1$ and $H_2$ in this expression. Obviously, if we take the alternating magnetic
f\/ield $H_1=-H_2=H$, then the total magnetization
is equal to zero, though the sublattice magnetizations remain non-trivial:
\[
m_{\rm odd}(H)=\left(1+\frac{1}{\pi}\int_0^\pi\frac{d\nu}
{\frac{\theta_4^2(\nu,q^2)}{\theta_1^2(\nu,q^2)}\cosh^2\beta H
+\sinh^2\beta H} \right)\tanh\beta H.
\]
However, after substituting $\beta H=h/(q^{-1}-q)$ and taking the limit
$\alpha\to 0$, we see that our zero temperature critical phenomenon disappears:
$m_{\rm odd}(h)= 1$ for $h>0$ and $m_{\rm odd}(h)= -1$ for $h<0$, similar
to the free spins system.

\section{Conclusion}

As shown in \cite{LS2}, soliton solutions of integrable hierarchies with
 the complex values of spectral variables are connected to the
intrinsic Coulomb gases on two dimensional lattices with some nontrivial dielectric
or conductor boundaries. In this picture,
the Coulomb interaction energy between two charges and their ef\/fective
images created by the boundary conditions plays the
role of the soliton phase shifts, the coordinates of charges
coincide with the spectral parameters of solitons, and
the external electrostatic f\/ield is expanded in some series
with the coef\/f\/icients playing the role of integrable hierarchy
times. This transparent relation serves as a clue for
building new Coulomb lattice gas models exactly solvable
at some f\/ixed temperatures. The latter temperatures are
also related to the random matrix models \cite{Gaudin,LS1,LS2},
but we do not discuss here applications of the described phase
transition within these interpretations.

Our phase transitions are of a rather simple nature.
In the lattice gas language, the transition in the $M=1$ periodic case
describes the situation when the lattice is f\/illed to its limit, i.e.\
the number of particles equals to the number of sites, and no more
particles can be added to the system. In the $M$-periodic case, such
transitions may happen separately, when each sublattice is f\/illed
completely one by one, or simultaneously, for an appropriate choice of parameters.
Their qualitative features can be found from the mean f\/ield theory.
There are several interesting questions which would be interesting to analyze
in the future, like inf\/luences of the hierarchy times on the thermodynamical
quantities, understanding of our systems beyond the ``solitonic" temperature
values, investigation of the higher correlation functions, and so on.

\subsection*{Acknowledgements}

The work of I.M. Loutsenko has been supported by
European Community grant MIFI-CT-2005-007323 and
V.P. Spiridonov is partially supported by the Russian
Foundation for Basic Research (grant no. 05-01-01086).
The authors are grateful to  V.B. Priezzhev for useful remarks.

\bigskip

{\it
I vaguely remember a tall man with glasses vigorously explaining to me
something during my poster presentation at the IGTMP
colloquium in Moscow in 1990. Probably that was Vadim~-- I~never asked him about that later on. We got acquainted at
the first SIDE meeting
near Montr\'eal in 1994 and had sufficiently long discussions
during his few days visit to CRM after that confe\-rence.
I remember telling him that by the work on separation of variables {\rm \cite{Kuz}},
which impressed me much, he closed to me that field, and it is necessary to
think about other directions of research.
We became closer during Vadim's stay at the CRM in 1994--1995.
It was very nice time from many points of view.
I visited his house and have known his family  during one of the
parties he was gathering. It appeared that he likes Russian
``bards" singing, which I was bond to as well. Once he even sent to me a web-link
to some new mp3-recordings coming from his native Saint-Petersburg.

 In June 1999, he chaired
my talk at the Hong Kong meeting on special functions, where I~reported
results of the paper {\rm \cite{spi-zhe:spectral}}.
He was interested by that much, and we discussed possible intersections
with integrable systems. I saw Vadim last time at the conference in
Edinburgh in September 2003, where he was the main organizer.
During the preparation of the corresponding proceedings,
I actively communicated with him, and it was clear that
he is extremely busy by all kinds of obligations.
In May 2005, I suggested to him to form a team in order to try to get
an INTAS grant. His first reaction was positive, but after he has
known the rules and procedures,
he rejected this idea by saying that there is too much bureaucracy and
he has too many other commitments for the next few months.
I totally agreed with his critics and accepted his excuse.
Later on he listed to me a number of other possibilities to get research
funding from the UK sources which sounded quite reasonable.
We exchanged by about ten e-mails with him over the May--October 2005
period, and it was devastating to know that he has passed away.
We have another deeply regrettable loss in the FSU scientific community,
which was possible, probably, to prevent in other circumstances.\\
%(Written by V.P. Spiridonov.)
\rightline{V.P. Spiridonov}

\bigskip

I have met Vadim first when I have been pursuing my PhD studies at
the Centre de Recherches Math\'ematiques in Montr\'eal.
At that time, Vadim was a postdoctoral fellow there.
I remember him as an open-heart person, frank and honest,
very enthusiastic and completely devoted to the problems
he did and had in mind. Always full of energy, he showed the
keenest interest for many questions of science and life.
To meet Vadim was very interesting to me.\\ %(Written by I.M. Loutsenko.)
\rightline{I.M. Loutsenko}
}

\pdfbookmark[1]{References}{ref}
\LastPageEnding


\begin{thebibliography}{99}

\footnotesize\itemsep=0pt

\bibitem{AS} Ablowitz M.J., Segur H., Solitons and the inverse
scattering transform, SIAM, Philadelphia, 1981.

\bibitem{aar:special}
Andrews  G.E., Askey R., Roy R.,
Special functions, {\it Encyclopedia of Math. Appl.},
Vol.~71, Cambridge Univ. Press, Cambridge, 1999.

\bibitem{bax} Baxter R.J., Exactly solved models in
statistical mechanics, Academic Press, London, 1982.

\bibitem{DJKM} Date E., Jimbo M., Kashiwara M., Miwa T.,
Transformation groups for soliton equations, in Nonlinear
Integrable Systems, World Scientif\/ic, Singapore, 1983, 41--119.

\bibitem{eva} Evans M. R., Phase transitions in one-dimensional nonequilibrium
systems, \href{http://arxiv.org/abs/cond-mat/0007293}{cond-mat/0007293}.

\bibitem{Gaudin}
 Gaudin M.,  Une famille \`a une param\`etre d'ensembles
unitaires,  {\it Nucl. Phys.} {\bf 85} (1966), 545--575.\\
 Gaudin M.,
Gaz coulombien discret \`a une dimension,
{\it J. Phys. (France)} {\bf 34} (1973), 511--522.

\bibitem{Hir} Hirota R., Exact solution of the Korteweg--de Vries
equation for multiple collisions of solitons,  {\it Phys. Rev. Lett.} {\bf 27}
(1971), 1192--1194.

\bibitem{Kuz} Kuznetsov V.B., Quadrics on real Riemannian spaces of constant
curvature: separation of variables and connection with Gaudin magnet,
 {\it J. Math. Phys.} {\bf 33} (1992), 3240--3254.

\bibitem{LS1} Loutsenko I.M., Spiridonov V.P., Self-similar potentials
and Ising models, {\it Pis'ma v ZhETF (JETP Letters)} {\bf 66} (1997), 747--753.\\
Loutsenko I.M., Spiridonov V.P.,
Spectral self-similarity, one-dimensional Ising chains and random matrices,
{\it Nucl. Phys.~B} {\bf 538} (1999), 731--758.%(3) [FS]

\bibitem{LS2} Loutsenko I.M., Spiridonov V.P., Soliton solutions of
integrable hierarchies and Coulomb plasmas, {\it  J. Stat. Phys.}
{\bf 99} (2000), 751--767, \href{http://arxiv.org/abs/cond-mat/9909308}{cond-mat/9909308}.

\bibitem{MS} Matveev V.B., Salle M.A.,
Darboux transformations and solitons,
Springer Series in Nonlinear Dynamics, Springer-Verlag, 1991.

\bibitem{spi:universal}
Spiridonov V.P., Universal superpositions of coherent states
and self-similar potentials, {\it Phys. Rev.~A} {\bf 52} (1995), 1909--1935,
\href{http://arxiv.org/abs/quant-ph/9601030}{quant-ph/9601030}.

\bibitem{spi-zhe:spectral} Spiridonov V.P., Zhedanov A.S.,
Spectral transformation chains and some new biorthogonal
rational functions, {\it Comm. Math. Phys.} {\bf 210} (2000), 49--83.


\end{thebibliography}
\end{document}